\documentstyle[preprint,aps,prb]{revtex}
\tightenlines

\def\cm-1{\hbox{cm$^{-1}$}}

\begin{document}
\draft
\title{Optical Conductivity of Manganites: Crossover from Jahn-Teller Small Polaron
to Coherent Transport in the Ferromagnetic State}
\author{M. Quijada,$^{(a)}$\cite{byline} J. \v {C}erne,$^{(a)}$ J. R. Simpson,$%
^{(b)} $ H. D. Drew,$^{(a,b)}$ K. H. Ahn,$^{(c)}$ A. J. Millis,$^{(c)}$ R.
Shreekala,$^{(b)}$ R. Ramesh,$^{(b)}$ M. Rajeswari,$^{(b)}$ and T. Venkatesan%
$^{(b)}$}
\address{$^{(a)}$Laboratory for Physical Sciences, College Park, Maryland 20740}
\address{$^{(b)}$Materials Research Laboratory, Department of Physics,\\
University of Maryland, College Park, Maryland 20742}
\address{$^{(c)}$Department of Physics and Astronomy, The Johns Hopkins \\
University, Baltimore, MD 21218}
\date{\today }
\maketitle

\begin{abstract}
We report on the optical properties of the hole-doped manganites Nd$_{0.7}$Sr%
$_{0.3}$MnO$_{3}$, La$_{0.7}$Ca$_{0.3}$MnO$_{3}$, and La$_{0.7}$Sr$_{0.3}$MnO%
$_{3}$. The low-energy optical conductivity in the paramagnetic-insulating
state of these materials is characterized by a broad maximum near 1 eV. This
feature shifts to lower energy and grows in optical oscillator strength as
the temperature is lowered into the ferromagnetic state. It remains
identifiable well below $T_{c}$ and transforms eventually into a Drude-like
response. This optical behavior and the activated transport in the
paramagnetic state of these materials are consistent with a Jahn-Teller
small polaron. The optical spectra and oscillator strength changes compare
well with models that include both double exchange and the dynamic
Jahn-Teller effect in the description of the electronic structure.
\end{abstract}


\preprint{}


\section{Introduction\newline
}

The discovery of colossal magneto-resistance (CMR) in hole-doped
ferromagnetic manganite materials of the form (Ln)$_{1-x}$(A)$_{x}$MnO$%
_{3-\delta }$, where Ln is a lanthanide and A is an alkaline-earth element,
has revived interest in this complex magnetic system.\cite{kuster1} These
systems crystallize in a pseudo-cubic perovskite structure. The
electronically active orbitals are the Mn d-orbitals and the number of
electrons per Mn is $4-x$. Hund's rule coupling implies that 3 electrons are
localized on the $t_{2g}$\ orbitals forming a core spin of $S_{c}=3/2$,
while the remaining $1-x$ electrons go into a band derived from the $e_{g}$\
orbitals. When the material is doped in the range of $0.2<x<0.5$, it
undergoes a phase transition between a paramagnetic insulator and a
ferromagnetic metal at temperature $T_{c}$.

The double-exchange model proposed by Zener\cite{zener} was the mechanism
first used to explain the paramagnetic to ferromagnetic phase transformation
in these conducting materials. In this model the hopping probability between
Mn sites for the electrons residing on the $e_{g}$\ orbitals is maximal when
the core spins are parallel and minimal when they are antiparallel. Hence,
the model provides a qualitative explanation for the metallic conduction
below $T_{c}$\ (spins well aligned), and insulating behavior above $T_{c}$\
(spins randomly aligned). However, recent theoretical as well as
experimental reports suggest that, within this model, the coupling between
the charge and spin is not enough to explain the high temperature insulating
phase. In particular, theoretical arguments\cite
{millisprl95,millisprl,millisprb,rother} indicate the need to include
effects that are related to the strong coupling between charge and the
lattice such as the Jahn-Teller (JT) effect associated with the Mn$^{3+}$
ions. In this picture a JT distortion lifts the degeneracy in the $e_{g}$
orbitals causing a splitting into lower and upper $e_{g}$ levels that is
static for $0<x<0.2$ and appears to be dynamic for $x>0.2$. Since the JT
distortion is associated with the Mn$^{3+}$ ions it can localize the $e_{g}$
electrons in the paramagnetic phase of the alloys leading to insulating
behavior. Whereas, the increased $e_{g\text{ }}$band width in the
ferromagnetic state quenches the JT effect and produces metallic conduction.
Optical absorption studies of the stoichiometric parent compound LaMnO$_{3}$%
\ give evidence for static JT distortions. Analysis of the optical
conductivity of LaMnO$_{3}$ within the local-spin-density approximation
suggests that the observed gap in the optical conductivity of about 1.0 eV
corresponds to the optical process of promoting a hole between the JT split $%
e_{g}$\ bands on the Mn$^{3+}$ ions.\cite{jung,hamada} Optical measurements
have also been reported for the series of compounds La$_{1-x}$Sr$_{x}$MnO$%
_{3}$.\cite{okimoto,tokura,arima} The results from optical reflectivity
studies of metallic samples ($x>0.2$) show large transfers of spectral
weight from high frequencies to low frequencies as the samples are cooled
from the paramagnetic state through $T_{c}$\ into the ferromagnetic metallic
state.\cite{okimoto,tokura} In these studies it is also concluded that a
simple double-exchange picture is not enough to explain the observed changes
in the optical spectral weight over such a large energy scale compared to $%
k_{B}T$ for all doping concentrations.\cite{okimoto,tokura} Other recent
experimental results also suggest the presence of strong lattice effects in
these materials: shifts in the IR phonon frequencies related to the Mn$-$O
bonds in La$_{0.7}$Ca$_{0.3}$MnO$_{3}$ near $T_{c}$ have been reported by
Kim $et.al.$\cite{kim}; anomalies in the local structure of the MnO$_{6}$
octahedron near $T_{c}$\ obtained from neutron scattering studies of La$%
_{1-x}$Ca$_{x}$MnO$_{3}$ have been interpreted in terms of Jahn-Teller
distortions\cite{billinge}; a magnetic field driven structural phase
transformation was observed\cite{asamitsu} in La$_{1-x}$Sr$_{x}$MnO$_{3}, $%
for $x=0.170$; thermopower measurements of La$_{0.7}$Ca$_{0.3}$MnO$_{3}$\
have also suggested the presence of small polaronic behavior above $T_{c}$\
that is found to disappear below the insulator-metal transition.\cite{jaime}

Localization of the $e_{g}$ electron on the Mn$^{3+}$ ions in the
paramagnetic state of the doped manganites due to Jahn-Teller distortions is
a self-trapping effect, {\it i.e.}, a small polaron.\cite{millisprb} If the
carriers are localized due to this electron-phonon coupling, there should be
an optical signature of the small polaron associated with photo-induced
hopping of the carriers, as has been reported in systems such as TiO$_{2}$.%
\cite{mahan,emin} In an earlier publication we reported evidence for the JT
small polaron in Nd$_{0.7}$Sr$_{0.3}$MnO$_{3}$.\cite{simon} In this paper we
present an extension of that earlier work by comparing the optical
conductivity derived from transmittance and reflectance measurements of La$%
_{0.7}$Sr$_{0.3}$MnO$_{3}$, La$_{0.7}$Ca$_{0.3}$MnO$_{3}$ and an oxygen
annealed Nd$_{0.7}$Sr$_{0.3}$MnO$_{3}$ thin film as a function of
temperature and for photon energies up to 5 eV. As in the previous work, the
optical conductivity on these new films show large shifts in spectral weight
from visible to infrared frequencies as the temperature is lowered through $%
T_{c}$, demonstrating broadband changes in electronic properties on an
energy scale several orders of magnitude larger than $k_{B}$$T_{c}$. In the
paramagnetic-insulating state of all three materials the optical
conductivity at low energies is characterized by a broad maximum near 1 eV
as we reported earlier for Nd$_{0.7}$Sr$_{0.3}$MnO$_{3}$\ and interpreted in
terms of the photon induced hopping of the Jahn-Teller small polaron.
Therefore these new experiments show that this feature is generic for the
paramagnetic state of the pseudo-cubic manganites. This spectral feature is
found to shift to lower energy and grow in optical oscillator strength as
the system enters into the ferromagnetic state. It gradually transforms into
a Drude-like response well below $T_{c}$ where the transport is metallic.
Nevertheless, the polaron peak can still be identified at temperatures
substantially below $T_{c}$.

At higher frequencies, the spectrum of \hbox{$\sigma_{1}(\omega)$}\
indicates the presence of a temperature dependent optical absorption feature
centered at 3 eV and a large spectral feature at 4 eV which still has a weak
temperature dependence. The 3 eV feature appears to be more prominent in the
paramagnetic state as is seen by looking at the difference in conductivity: $%
\Delta \sigma _{1}=\sigma _{1}(T)-\sigma _{1}(10K)$. The energy position of
this feature suggests that it involves transitions between the Hund's rule
spin-split $e_{g}$ derived bands. The 4 eV feature is most likely related to
a similar feature that is seen in the undoped materials where it has been
identified as a charge transfer transition between the O$_{2p}$ and the Mn$%
_{d}$ derived bands.\cite{arima}

We also present an analysis of the magnitude and temperature dependence of
the spectral weight $\int d\omega \sigma _{1}(\omega )$ and of its relation
to the spin wave stiffness at $T=0$.\cite{millisprl95} This analysis extends
and corrects an earlier treatment by including the finite Hund's rule
splitting and a more realistic treatment of the $e_{g\text{ }}$bands.\cite
{millisprl95,millisprl,millisprb,rother}

\section{Experimental\newline
}

\subsection{Sample characterization}

The samples used in this study were thin films of Nd$_{0.7}$Sr$_{0.3}$MnO$%
_{3}$, La$_{0.7}$Sr$_{0.3}$MnO$_{3}$, and La$_{0.7}$Ca$_{0.3}$MnO$_{3}$\
grown on LaAlO$_{3}$ substrates by pulsed laser deposition in an N$_{2}$O
atmosphere\cite{2}(in what follows we will refer to these samples as NSMO,
LSMO, and LCMO respectively). The films are epitaxial as revealed by x-ray
diffraction, and show 3 MeV He$^{+}$ ion Rutherford backscattering
channeling spectra with a minimum yield of 3.8\%, indicating a high degree
of crystallinity. Additional characterization of these samples was done by
performing resistivity and ferromagnetic resonance measurements.\cite
{lofland,sam}

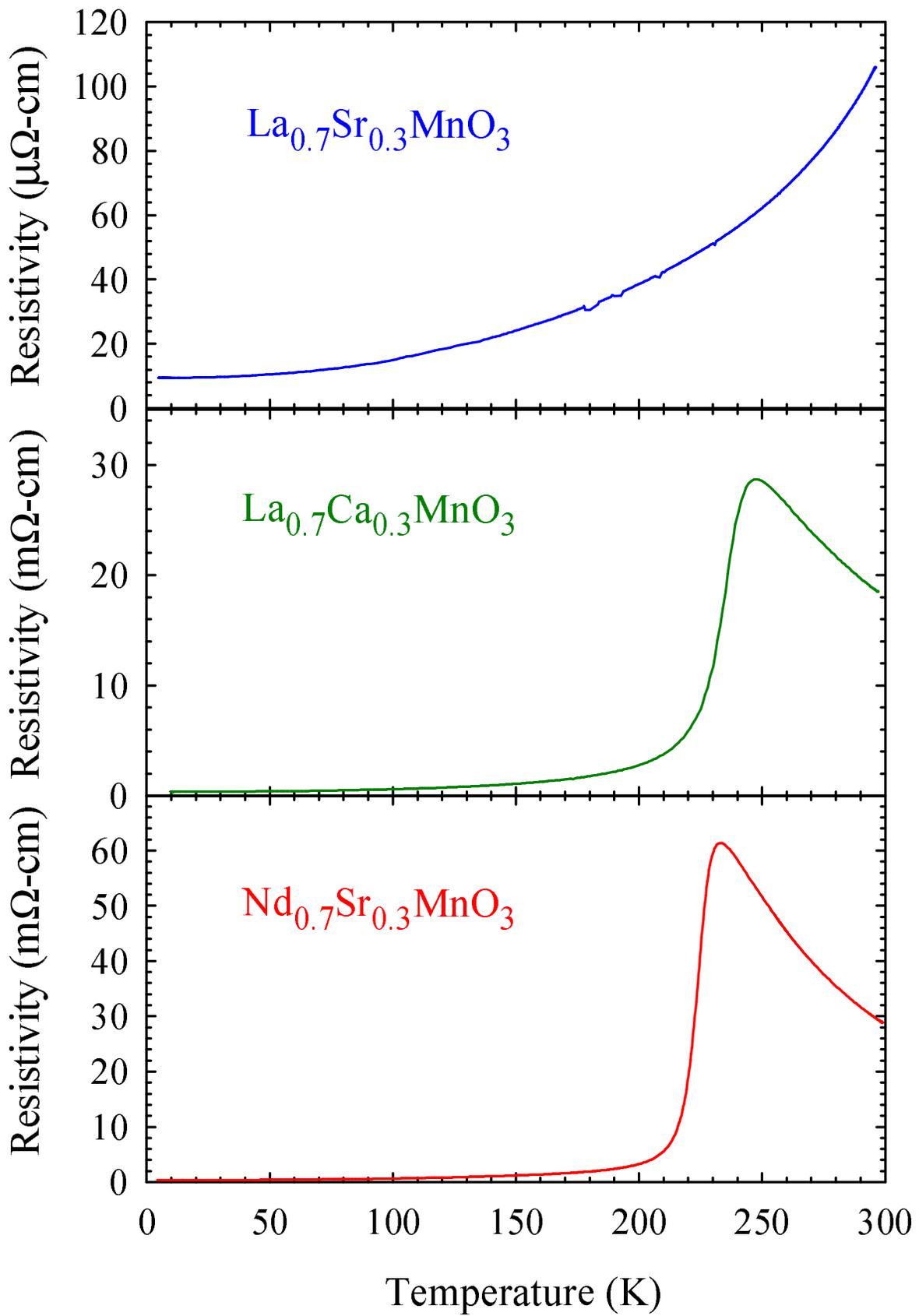
\begin{figure}[tbp]
\caption{Temperature dependence of the resistivity for the three samples.
The metal to insulator transition is apparent for the NSMO and LCMO.}
\label{fig1}
\end{figure}

Fig.~\ref{fig1} shows the temperature dependence of the resistivity for the
three samples that we studied using standard four-probe measurements. The
temperature at which the resistivity peaks, $T_{p}$, is 235 K for NSMO, 250
K for LCMO and 360 K for LSMO. These values are close to the Curie
temperature $T_{c}$\ in these samples. In addition to the observed
progression in $T_{c}$, there is also a variation in the low-temperature
value of the resistivity $\rho$. The values for $\rho $ at $T=10$ K are 350 $%
\mu \Omega -{\rm cm}$ for NSMO, 300 $\mu \Omega -{\rm cm}$ for LCMO, and 10 $%
\mu \Omega -{\rm cm}$ for the LSMO sample. To the best of our knowledge,
these are the lowest values for the resistivities ever reported for these
materials indicative of the high quality of these thin film samples. The
observed low temperature resistivities correspond to conductivities greater
than the Mott minimum conductivity ($\sigma _{Mott}=0.656\frac{e^{2}}{h}%
n^{1/3}\approx 500\,\Omega ^{-1}{\rm cm}^{-1}$ if the carrier density is
determined by the divalent alkaline earth substitution $x/\Omega _{0}$,
where $\Omega _{0}$ is the unit cell volume). It is noteworthy that the low
temperature resistivities previously reported in the literature for bulk and
film samples generally correspond to conductivities below the Mott minimum
conductivity which suggests that those samples were not homogeneous. Our
results imply that previous reports do not reflect intrinsic behavior,
perhaps because of sample inhomogeneity. We suggest that proposals of
``super-unitary'' scattering\cite{coey,rama} be re-examined as the intrinsic
behavior becomes better determined.

\subsection{Optical techniques}

Transmittance and reflectance measurements were performed using a Fourier
transform spectrometer (BOMEM DA3) to cover the investigated regions of 5
meV to 25 meV and 0.20 eV to 5 eV. These are the frequency ranges for which
the LaAlO$_{3}$ substrate is transparent. Measurements in this wide range of
frequencies were accomplished by using different combinations of sources,
beamsplitters and detectors. The uncertainty in the absolute transmittance
obtained from the reproducibility in the different spectral ranges is $\pm 2$%
\%. Likewise, the uncertainty in the absolute reflectance spectra is
estimated at $\pm 3$\%. Temperature dependent measurements were made
possible by mounting the samples on the cold finger of a continuous flow
cryostat with room temperature windows for optical access. The cryostat unit
was placed inside the sample compartment of the spectrometer. The
temperature of the sample was stabilized by using a temperature controller
with a calibrated Si diode sensor mounted near the sample and a heating
element attached to the cold finger of the cryostat.

At far infrared frequencies the real part of the optical conductivity, $%
\sigma _{1}$, is determined from the observed transmittance given by 
\begin{equation}
\frac{T_{FS}}{T_{S}}=\left[ \left( 1+Z\sigma _{1}\right) ^{2}+\left( Z\sigma
_{2}\right) ^{2}\right] ^{-1},  \label{tfst}
\end{equation}
in the thin film limit $d/\delta <<1$, which is well satisfied in our
samples (here $\delta =\lambda /4\pi \kappa ,$ is the skin depth and $d$ is
the film thickness). $T_{FS}$ and $T_{s}$ are the transmittance of the
film/substrate and substrate respectively, $Z=\frac{Z_{0}d}{n+1}$, where $%
Z_{0}$ is the free space impedance, and $n$ the refractive index of the
substrate. Inverting Eq.\ (\ref{tfst}) we can write 
\begin{equation}
\sigma _{1}=Z^{-1}\left[ \sqrt{\frac{T_{S}}{T_{FS}}-\left( Z\sigma
_{2}\right) ^{2}}-1\right] \leq Z^{-1}\left[ \sqrt{\frac{T_{S}}{T_{FS}}}%
-1\right]  \label{trans}
\end{equation}
The final inequality becomes an equality when $\sigma _{2}/\sigma _{1}<<1$.
Terahertz measurements on a LSMO sample\cite{andrea} confirm that $\sigma
_{2}/\sigma _{1}<<1$ is a valid approximation in the regime of interest $%
0<\omega <100~{\rm cm}^{-1}$.

At higher frequencies, the optical properties of the samples in the
infrared-UV range are derived from the measured transmittance and
reflectance by numerically inverting the Fresnel formulae for a thin film on
a weakly absorbing thick substrate. The interference (etalon) effects from
the substrate are averaged by using low spectral resolution in the
measurement and by averaging over the interference fringes in the analysis.%
\cite{heavens} This procedure yields the index of refraction and extinction
coefficient of the thin-film material, $n_{f}$ and $\kappa _{f}$
respectively, using $n$ and $\kappa $ for the substrate. The results of $%
n_{f}$ and $\kappa _{f}$ are then used to derive the other optical constants
such as the dielectric function, $\epsilon (\omega )$, or the optical
conductivity \hbox{$\sigma(\omega)$}.\cite{wooten} Since this technique does
not rely on Kramers-Kronig analysis it permits reliable measurement of the
optical conductivity up to the high frequency cut off of the optical
equipment $(5$ eV$)$.

This analysis assumes that the index of refraction, $n$, and $\kappa $ of
the substrate are known quantities. In all the substrate samples that we
measured, we found the index to be roughly constant, $n\simeq 2.0$, and
sample independent in the 0.2$-$5 eV range.\cite{zhang} Although the values
for the extinction coefficient were small in this same range ($%
10^{-3}-10^{-4}$), we found some variation in $\kappa $ for different
samples of LaAlO$_{3}$ especially near the cutoff frequency of 0.2 eV. The
differences are sufficiently large that it is necessary to use the values
for $n$ and $\kappa $ for the substrate on which the film was grown in the
analysis since the effect of $\kappa $ in the substrate is similar to the
effect of $\sigma _{1}$ in the sample ( $\sim 10$\%). Therefore, after the
film was measured, it was removed from the substrate and the transmittance
and reflectance of the bare substrate were measured. This procedure ensured
that the proper values of $n$ and $\kappa $ for the substrate were used in
the final inversion of the data.

\section{Results: Optical spectra \newline
}

The optical conductivity in the far infrared was estimated as described in
section II-B. Because of the narrow range of the transmission window of the
LaAlO$_{3}$ substrates and the featureless transmissivity from 20 $-$ 100 cm$%
^{-1}$ these data provide only a low frequency data point to our broad band
conductivity spectra shown in Fig.~\ref{fig2}. It is interesting to compare
this far infrared conductivity with the DC conductivity of the same films.
Fig.~\ref{fig3} displays a comparison of the measured DC conductivity and
the derived conductivities at 20 cm$^{-1}$ using Eq.\ (\ref{trans}). We
noticed a reasonable agreement between the AC and DC values for LCMO and
NSMO samples. However, a striking disagreement is found for the LSMO sample.
The measured DC conductivity in this sample is roughly 10 times larger than
the AC values obtained at 20 cm$^{-1}$. Similar results have been observed
in another high conductivity LSMO film. However, typical LSMO films have a
resistivity of around 100 $\mu \Omega -{\rm cm}$ in better agreement with
the AC values. The observed discrepancy for the high conductivity films is
particularly significant when it is recalled that the far infrared AC
conductivity derived from Eq.\ (\ref{trans}) is an upper limit of the true
value. Therefore, the discrepancy is not a consequence of granularity in the
film since this effect would lead to an effective DC resistivity larger than
the intrinsic value due to the higher resistivity of the grain boundaries.
We believe that the observed behavior may be explained in terms of an
anisotropic conductivity arising from strain effects in the films. Such
effects may be expected in materials with a strong coupling between the
electrons and the lattice as we are reporting in this paper. Further studies
of these effects are in progress.

The results of the real part of the optical conductivity, %
\hbox{$\sigma_{1}(\omega)$}, for the three samples measured at different
temperatures are shown in Fig.~\ref{fig2}. The symbols near zero frequency
are obtained from the far-infrared transmittance measurements as described
above. The IR conductivity is seen to extrapolate reasonably to the far
infrared value except for a downturn near the low frequency IR cutoff. This
effect is believed to be related to the difficulties in characterizing the
differences in the extinction coefficient of the substrates near the IR
cutoff as discussed earlier. Therefore we do not trust the data below about
0.3 eV. In the results shown in Fig.~\ref{fig2}, we find that near and above 
$T_{c}$, the optical conductivity for the three samples is dominated by a
broad maximum near 1 eV with a peak value of roughly 600$-$700 $\Omega
^{-1}- $cm$^{-1}$, only slightly higher than the Mott conductivity. We also
notice that the feature near 1 eV evolves in temperature in very much the
same way for the different alloys. These data indicate that this feature is
universal in the hole doped pseudo-cubic manganites. In all the samples, the
broad maximum in the conductivity spectrum above $T_{c}$ moves to lower
frequencies and grows in oscillator strength as temperature is decreased
through $T_{c}$. The peak structure remains identifiable well below $T_{c}$,
but as the temperature is lowered further into the metallic range the low
frequency part fills in and eventually \hbox{$\sigma_{1}(\omega)$} increases
with decreasing frequency for $T<<T_{c}$. This indicates a Drude response
and coherent conduction at low temperatures and low frequencies in all the
samples. The results for the new oxygen annealed Nd$_{0.7}$Sr$_{0.3}$MnO$%
_{3} $\ sample differ from the results on an unannealed sample, reported
previously, which did not show a Drude-like response even at the lowest
temperature, although the $T>T_{c}$ behavior was similar to that observed
here. The earlier sample had a lower $T_{c}$\ and a much higher DC
resistivity $\left( \sigma _{DC}<\sigma _{Mott}\right) $ than the present
sample even at the lowest $T$ and similar to values generally reported in
the literature on NSMO. Clearly, the different oxygen treatment of the two
samples produced different behaviors in their DC transport and optical
properties.

Throughout the infrared the reflectance and transmittance of the films is
dominated by $\sigma _{1}$ and is relatively insensitive to $\epsilon _{1}$.
Therefore our results $\epsilon _{1}$ are less reliable. Nevertheless, we
find that $\epsilon _{1}$ remains positive down to at least 0.5 eV despite
the apparent free-carrier response of \hbox{$\sigma_{1}(\omega)$} at low
temperature, which would be expected to produce a negative $\epsilon _{1}$
characteristic of a metal. The observed behavior is not completely
unexpected since interband transitions between the two $e_{g\text{ }}$bands
are expected in this spectral range. Nevertheless, this behavior indicates
that the spectral weight of the coherent component of the conductivity is
only a small fraction of the total spectral weight in all these samples as
has been reported in Refs.~[\onlinecite{okimoto,tokura}].

The other major spectral feature in the optical data is the strong peak in
the conductivity near 4 eV. The noisy appearance of the conductivity data
near the peak of this feature is a consequence of the very small
transmittance of the films due to the strong absorption at the peak. Also,
this feature appears noticeably weaker in Nd$_{0.7}$Sr$_{0.3}$MnO$_{3}$.

\section{Discussion\newline
}

\subsection{Overview\newline
}

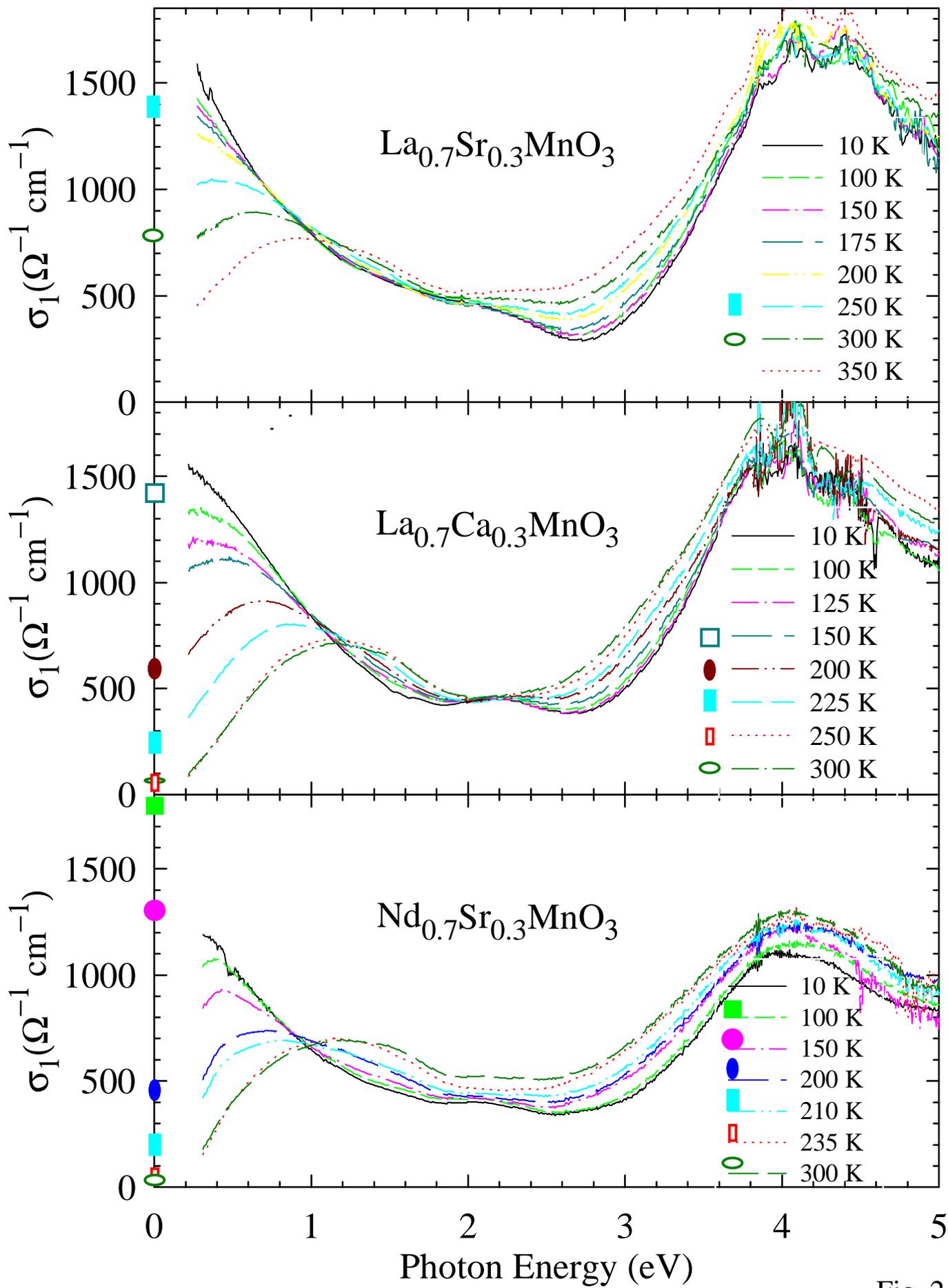
\begin{figure}[tbp]
\caption{Frequency dependence of the real part of the optical conductivity, $%
\sigma_1$, for the three samples at different temperatures.}
\label{fig2}
\end{figure}

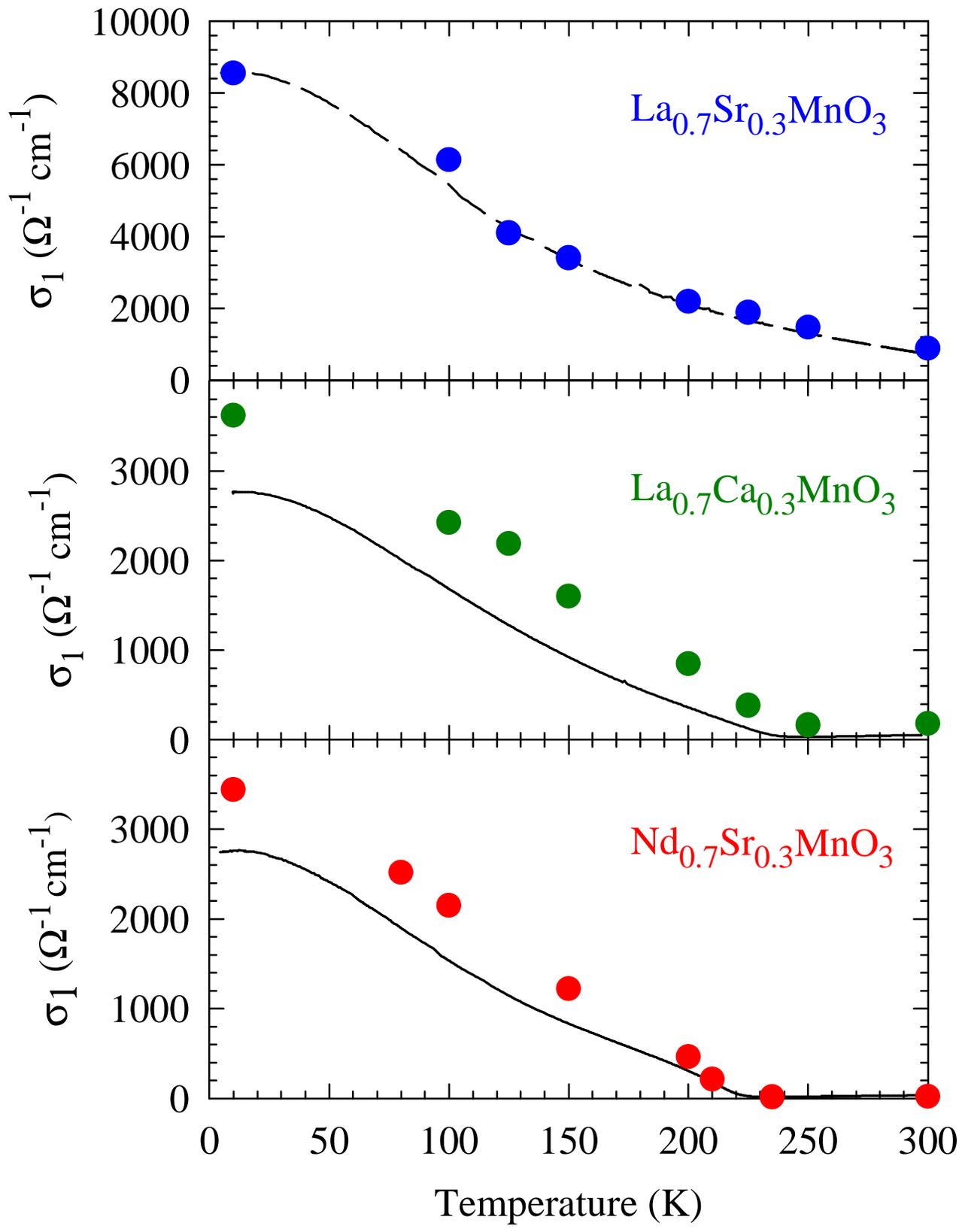
\begin{figure}[tbp]
\caption{Comparison of the temperature dependence of the dc (solid lines)
and ac (circles) conductivities at 20 cm$^{-1}$ derived from Eq. 2.2. The dc
conductivity for the La$_{0.7}$Sr$_{0.3}$MnO$_3$ sample (the dashed curve)
has been divided by 10 in order to fit the data in the same scale.}
\label{fig3}
\end{figure}

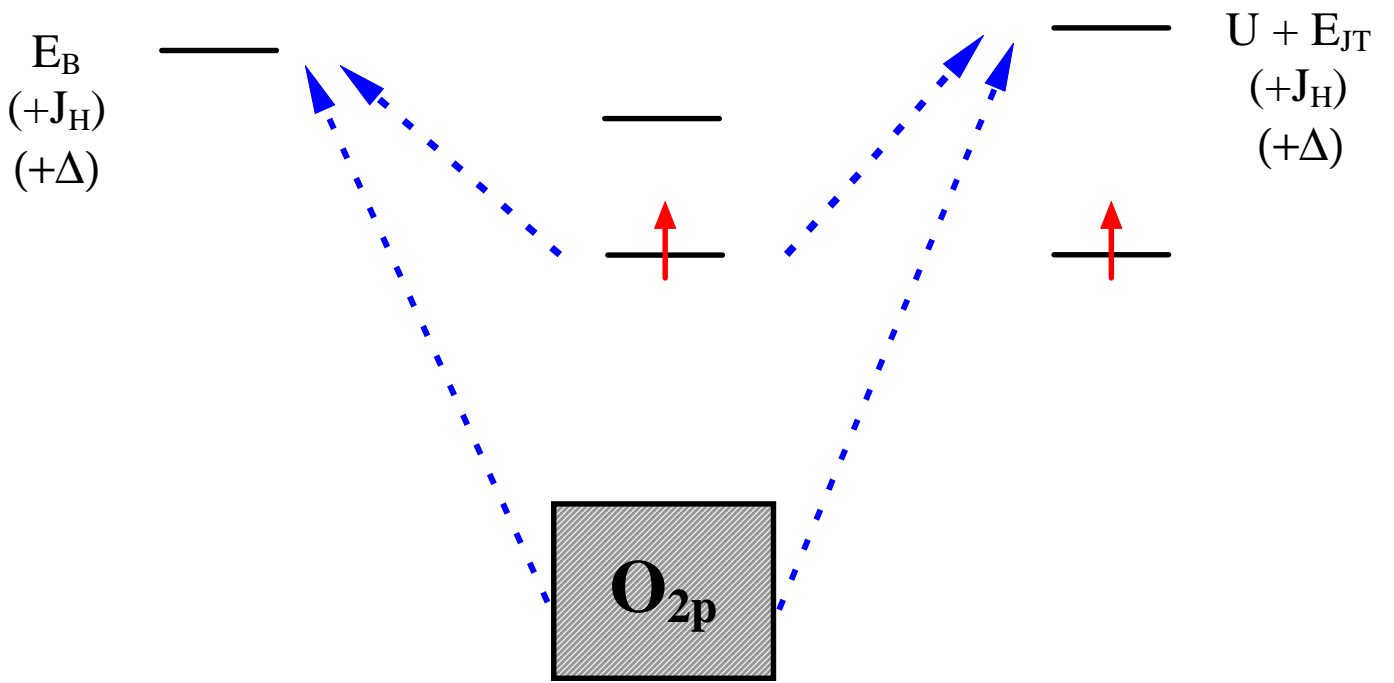
\begin{figure}[tbp]
\caption{States and energies involved in optical transitions. Central
portion shows initial states left and right panels show allowed optical
transitions and final state energies. The initial states in the central
panel are: a filled oxygen p-band and a Mn site with one $e_g$ electron. The
two fold degeneracy of the $e_g$ levels is assumed split by a local
Jahn-Teller distortion. The level splitting (not indicated on the figure) is 
$E_{JT}$. Right hand portion indicates optical transition in which charge is
moved from an O or Mn site to a Mn site which is already singly occupied.
The final state energy is the sum of the Jahn-Teller splitting $E_{JT}$, the 
$e_g-e_g$ Coulomb repulsion U, the charge transfer energy D (if the electron
came from an O site) and the Hund's energy $J_H$ (if the transferred
electron has spin antiparallel to the core spin). The left hand portion
indicates an optical transition to a Mn which initially has no $e_g$
electrons. In this case the $e_g$ levels are degenerate and the final state
energy is shifted by the breathing distortion energy $E_B$ (and $\Delta$ and 
$J_H$, if appropriate).}
\label{fig5}
\end{figure}

In this section we discuss in more detail several of the striking features
of the optical conductivity data shown in Fig.~\ref{fig2}. There are three
main peaks: at $\sim 1$ eV, 3 eV, and 4 eV. A qualitative picture of the
states and optical transitions is shown in Fig.~\ref{fig5}. The important
orbitals are the Mn $e_{g}$ and O$_{2p}$ levels. Optical absorption is
controlled by the electric dipole matrix element and allowed transitions
involve motion of a charge either from one Mn-site to another or from an O
to a Mn. Ref.~[\onlinecite{korean98}] argues that transitions from one $%
e_{g} $ orbital to another one {\it on the same site} appear in the optical
spectrum, but we believe these transitions have negligible oscillator
strength because both the initial and the final states have d-symmetry with
respect to the same origin.

There are several important energies. One is the $e_{g}-O_{2p}$ charge
transfer energy $\Delta$. Another is the Jahn-Teller energy $E_{JT}$: if a
Mn site has one $e_{g}$ electron then a local even-parity lattice distortion
which breaks cubic symmetry may occur; this splits the $e_{g}$ levels by an
amount $E_{JT}$. A third is the breathing distortion energy $E_{B}$: at an
Mn-site with no $e_{g}$ electrons a symmetric ``breathing'' distortion of
the surrounding oxygens may occur: if (as occurs in an optical transition)
an $e_{g}$ electron is added on a timescale which is fast compared to a
phonon frequency (so the lattice does not have time to relax) the final
state is shifted up in energy by $E_{B}$. In LCMO with $x=0.5$, the low
temperature phase is charge ordered and both breathing and Jahn-Teller
distortions occur.\cite{Andywillget} The amplitudes of these distortions are
approximately equal, suggesting $E_{B}\sim E_{JT}$. A fourth energy is the
``Hubbard $U$'' repulsion\cite{chainani} between two $e_{g}$ electrons on
the same site. As will be discussed below, available evidence suggests that
the effective $U$ describing the low ($\omega <4$ eV) energy physics of the $%
e_{g}$ band is weak. A final energy is the Hund's coupling $J_{H}$: each Mn
has a $S=3/2$ t$_{2g}$ core spin (not shown in Fig.~\ref{fig5}); any $e_{g}$
electrons may have spin parallel or antiparallel to the core spin; the
``antiparallel'' state is higher by an energy $J_{H}$.

In the remainder of this section we shall argue that the 1 eV feature
involves $e_g-e_g$ transitions within the parallel spin manifold; the 3 eV
feature is due to $e_g-e_g$ transitions in which the final state is
antiparallel and the 4 eV feature is the $e_g-$O$_{2p}$ transition.

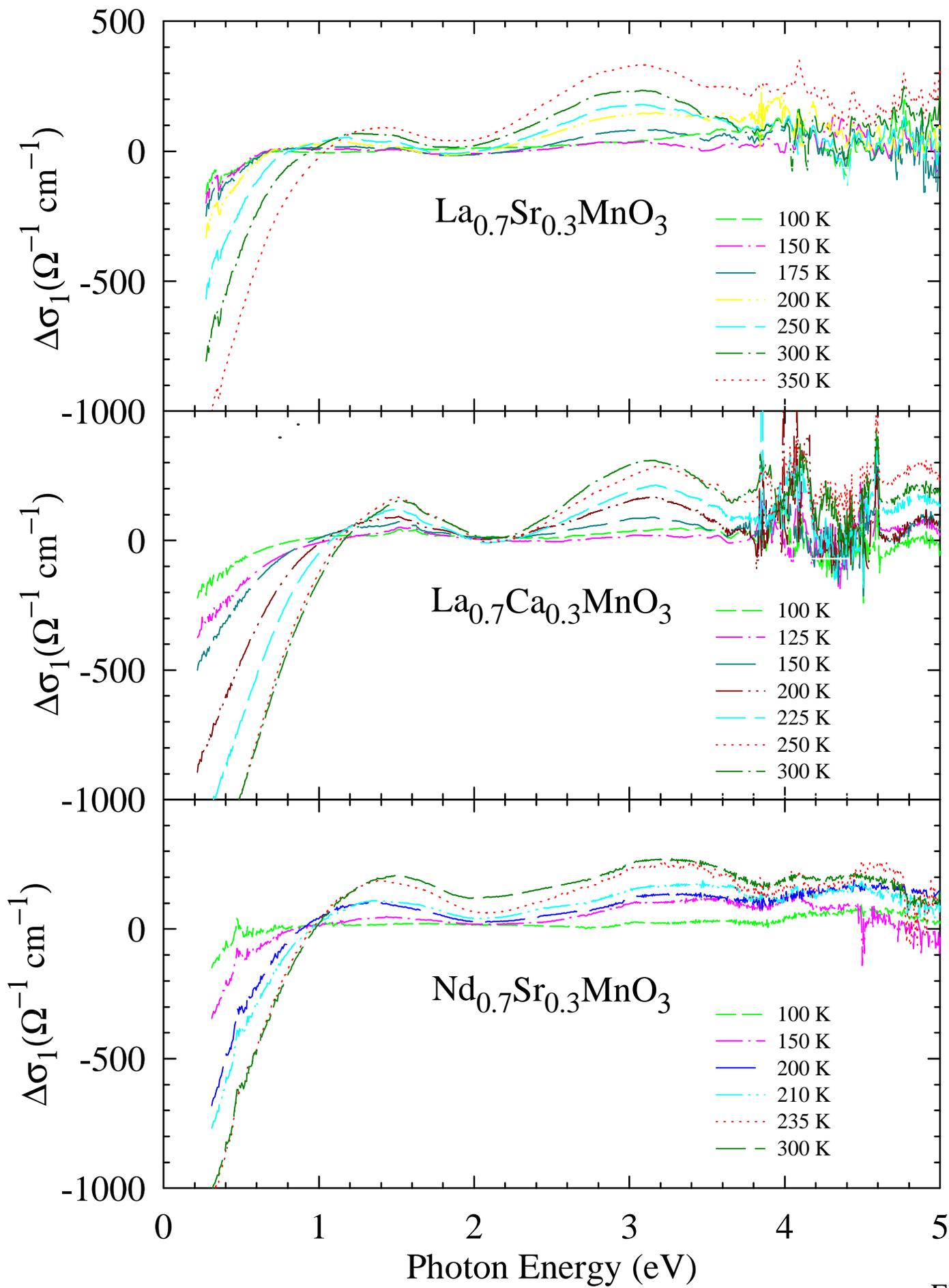
\begin{figure}[tbp]
\caption{Difference in optical conductivity $\Delta \sigma_1 =
\sigma_1(T)-\sigma_1(10K)$ for the three samples. Notice the feature at 3.0
eV that shows up more prominently in the paramagnetic state.}
\label{fig6}
\end{figure}

The most striking feature of the data in this energy range is the broad peak
centered at $\omega \approx 1$ eV ($1.3\sim $ eV in the LCMO and NSMO
samples, and perhaps at a somewhat lower energy in LSMO), which is evident
in the paramagnetic phase in all samples and which, as $T$ is decreased
below the magnetic $T_{c}$, loses intensity and shifts to lower frequency,
eventually evolving into the Drude-like response observed at very low
temperatures. This feature was first observed by Okimoto $et.al.$\cite
{okimoto} and was interpreted by them as the ``parallel'' to
``antiparallel'' transition, implying a $J_{H}\sim 1.5$ eV. However, as
noted by Millis $et.al.$\cite{millisprl} the temperature dependence of the
spectral weight of the 1 eV feature is inconsistent with this
interpretation. Because the electric dipole matrix element preserves spin,
the ``antiparallel'' final state is inaccessible if the material is in the
fully polarized ferromagnetic ground state, so the oscillator strength in
the ``antiparallel'' transition should decrease as $T$ is lowered below $%
T_{c}$. However, as seen in Fig.~\ref{fig2} and Fig.~\ref{fig6}, the
intensity in the peak feature grows as $T$ is decreased below $T_{c}$, even
as the feature broadens and shifts to lower energies, inconsistent with the
``antiparallel'' interpretation. We, therefore, believe it involves parallel
spin $e_{g}-e_{g}$ transitions only. However, the temperature dependence of $%
\sigma (\omega )$ near $\omega =3$ eV is consistent with transitions to an $%
e_{g}$ ``antiparallel'' final state. As can clearly be seen in Fig.~\ref
{fig6}, which displays the difference [$\sigma _{1}(\omega ,T)-\sigma
(\omega ,T=10{\rm K})$], a peak at $\omega \approx 3$ eV grows in strength
as the temperature is raised and the core spins are disordered.

Returning now to the 1 eV feature, we ascribe it to $e_{g}-e_{g}$
transitions. As will be discussed in detail in the next section, the
oscillator strength is consistent with that expected for dipole allowed d$-$%
d charge transfer transitions between Mn ions on different sites. The
insulating nature of the paramagnetic phase and the peak-like shape of the
absorption suggests that it arises from excitation of carriers out of bound
states. One possibility is localization due to disorder, as suggested by
Ref.~[\onlinecite{houstonPRL}]. We think this is unlikely because of the
universal appearance of the feature in manganites of widely varying dc
conductivities.\cite{jrcat} As noted by Jung $et.al.$\cite{korean98} the
characteristic energy of the absorption feature in the paramagnetic state of
the doped materials is similar to that observed in insulating LaMnO$_{3}$,
where a $\sim 1.5$ eV gap appears in the $e_{g}$ manifold due to the
presence of a long range ordered Jahn-Teller distortion. In LaMnO$_{3}$ this
feature is believed to be due to the $e_{g}-e_{g}$ transition shown in the
right-hand panel of Fig.~\ref{fig5}; the gap of 1.5 eV corresponds to a
band-center to band-center energy difference of about 2.5 eV and we identify
this energy as $E_{JT}+U$ for LaMnO$_{3}$.

Doped materials in the paramagnetic phase were predicted \cite{millisprl95}
to have lattice distortions similar to those occurring in LaMnO$_3$ (but
without the long ranged order) and these distortions (with amplitude about
70\% of those in LaMnO$_3$) have recently been directly observed in neutron
PDF\cite{billinge} and EXAFS\cite{booth98} experiments. We, therefore,
interpret the 1 eV feature observed at $T>T_c$ in our doped samples as being
due to the excitation of a carrier out of a bound state produced by a strong
local lattice distortion (and into a final state also modified by the
lattice distortions). Of course, in doped materials the $e_g$ electron
concentration is less than 1 per site, so transitions such as those shown in
the left panel of Fig.~\ref{fig5} are also possible. The fact that only one
(broad) feature is observed suggests that the energy shift due to the
breathing distortion, $E_B$, is comparable to the Jahn-Teller plus Coulomb
energy $E_{JT}+U$.

As the temperature is decreased below $T_{c}$, the 1 eV feature grows in
intensity and broadens, and eventually evolves into the Drude-like response
observed at very low temperatures. This behavior corresponds to the collapse
of the Jahn-Teller small polaron as the system goes into the ferromagnetic
state. It can be understood in terms of a model proposed by Millis, Mueller
and Shraiman\cite{millisprl95} that incorporates double exchange and dynamic
JT effects in the system. In this model, the behavior of the system is
controlled by the dimensionless effective coupling constant $\lambda $
defined as 
\begin{equation}
\lambda ={\frac{{E_{JT}}}{{t\langle \cos (\theta _{ij}/2)\rangle }}},
\end{equation}
where $t$ is the hopping probability, and $\theta _{ij}$ is the relative
angle between neighboring spins. The temperature dependence is controlled by
the $\langle {\rm cos}(\theta _{ij}/2)\rangle $ factor, which goes from 2/3
in the paramagnetic state to 1 in the ferromagnetic state. Within this model
a qualitative description of the data shown in Fig.~\ref{fig2} is as
follows: In the paramagnetic state $\lambda $ is larger than $\lambda _{c}$,
the critical value for the formation of a small polaron. As the temperature
is decreased, the spins become aligned and $\lambda $ decreases becoming
smaller than $\lambda _{c}$ leading to a collapse of the JT small polaron
and coherent Drude-like conduction at low temperatures. The optical
conductivity calculated within this model show shifts in oscillator strength
and linewidth as function of $\lambda $ that compare well to the
experimental results shown in Fig.~\ref{fig2} below 2 eV.\cite{millisprb}
The behavior of the resistivities, the Curie temperatures and the optical
properties of these materials indicate that LSMO, LCMO, and NSMO have a
progressively increasing JT coupling $\lambda $. We emphasize that in all
the samples the JT small polaron feature remains for intermediate
temperatures below $T_{c}$. This is consistent with thermopower and EXAFS
measurements that also suggest evidence for small polaronic behavior in the
LCMO near $T_{c}$.\cite{jaime,booth98} The downward shift in the polaron
feature and the onset of metallic conductivity indicate a gradual transition
from a small polaron to a large polaron and a corresponding gradual growth
of the strength of the coherent conductivity.

Another aspect of the data shown in Fig.~\ref{fig2} is a large
redistribution of spectral weight from optical transitions occurring above 2
eV to below the 1 eV feature discussed above. The temperature dependence of %
\hbox{$\sigma_{1}(\omega)$}\ is seen to extend up to the upper limit of our
measurements (5 eV). In this section we discuss the origins of the two major
spectral features in this range. The first one is the strong absorption at $%
\omega \simeq 4.0$ eV. The assignment of the optical process involved in
this transition is made by comparing it with the result of the undoped
parent compound LaMnO$_{3}$. In this material, a similar peak in the optical
conductivity has been observed at this frequency and assigned to a charge
transfer transition between the O$_{2p}$ and the $e_{g}$ derived bands.\cite
{jung,arima} The $T$-dependence of the 4 eV feature does not, however,
account for the missing spectral weight when the sample is warmed above $%
T_{c}$ as can be seen in Fig.~\ref{fig6}. To fully account for all the
low-frequency oscillator strength in the ferromagnetic state it is clear
that it will be necessary to include contributions up to and beyond the 5 eV
limit of our measurements. The temperature dependence of the spectral weight
will be discussed further in the next section.

\subsection{Comparison to theory: magnitude and temperature dependence of
optical spectral weight \newline
}

This section presents a more quantitative comparison of data to theory. The
main results are: (i) the $e_{g}$ electron kinetic energy is largest (and
effective electron-phonon interaction weakest) in LSMO, while the kinetic
energy is smallest (and electron-phonon interaction strongest) in NSMO, with
LCMO being intermediate. (ii) In all compounds at lowest temperature the $%
e_{g}$ kinetic energy is greater than or approximately equal to the band
theory value, suggesting that at this temperature the effects of the
electron-phonon and electron-electron interaction are relatively weak. (iii)
The change in $e_{g}$ kinetic energy between lowest temperature and $T_{c}$
is too large to be consistent with models involving only double exchange,
but is consistent with theories based on the interplay of double exchange
and electron-phonon coupling. (iv) There is a reasonable consistency between
the optical estimates of the electron kinetic energy and the observed low
temperature spin wave stiffness (our analysis here generalizes and corrects
statements made in Ref.~[\onlinecite{millisprl95}]). (v) The magnitude of
the change in optical absorption near $\omega \approx 3$ eV is
quantitatively consistent with the interpretation, given in the previous
section, of the 3 eV peak as transitions to the ``antiparallel'' final state.

Because calculations of $\sigma (\omega ,T)$ including both a realistic
treatment of the $e_{g}$ band structure and the effects of the
electron-phonon interaction are not available, we focus the discussion on
the spectral weight $S(\omega )$ defined by 
\begin{equation}
S(\omega )=\frac{2}{\pi }\int_{0}^{\omega }d\omega \sigma _{1}(\omega ).
\label{three}
\end{equation}
It is sometimes convenient to define an effective number of electrons per
unit cell {\rm N}$_{eff}(\omega )$ and a kinetic energy {\rm K}$(\omega $)
via 
\begin{equation}
{\rm N}_{eff}(\omega )={\frac{V_{cell}m}{e^{2}\hbar ^{2}}}S(\omega ),
\label{two1}
\end{equation}
\begin{equation}
{\rm K}(\omega )={\frac{a_{0}}{e^{2}}}S(\omega ),  \label{K}
\end{equation}
where $V_{cell}$ is the unit cell volume and $a_{0}=V_{cell}^{1/3}$. If the
integral in Eq.\ (\ref{three}) is carried out to $\omega =\infty $ the
familiar f-sum rule implies one obtains {\rm N}$_{eff}(\omega =\infty )={\rm %
N}$, the total number, core plus valence of electrons in the unit cell. $%
S(\omega <\infty )$ gives a partial sum rule which may give information
about low-lying states of interest. For the manganites the low-lying states
of interest are the $e_{g}$ electrons; in the analysis which follows we will
first present some theoretical results, then determine an $\omega $ such
that $S(\omega )$ is a good estimate of the spectral weight in the $%
e_{g}-e_{g}$ transitions and finally use the magnitude and temperature
dependence of the determined spectral weight to infer the results listed
above.

We begin with theory. We assume the physics of interest is determined by
carriers hopping between Mn $e_{g}$-symmetry $d$-levels on sites of a simple
cubic lattice of lattice constant $a_{0}$ and interacting with each other,
with the lattice, and with core spins of magnitude $S_{c}=3/2$. The
Hamiltonian is 
\begin{equation}
{\hat{H}}=-{\frac{1}{{2}}}\sum_{%
{\scriptstyle i{\vec{\delta}} \atop \scriptstyle ab}%
}t_{\vec{\delta}}^{ab}\left[ e^{i\vec{A}\cdot \vec{\delta}}d_{ia\alpha
}^{\dag }d_{i+\vec{\delta}b\alpha }+h.c.\right] -J_{H}\sum_{%
{\scriptstyle ia \atop \scriptstyle \alpha \beta }%
}\vec{S}_{ci}\cdot d_{ia\alpha }^{\dag }{\vec{\sigma}_{\alpha \beta }}%
d_{ia\beta }+H_{INT}.  \label{hbar}
\end{equation}
Here $d_{ia\alpha }^{\dag }$ creates an electron with spin $\alpha $ in $%
e_{g}$ orbital $a$ on site $i$, $J_{H}$ is the Hund's coupling between the
itinerant electrons and the core spins, $H_{INT}$ represents the other
interactions and $t_{\vec{\delta}}^{ab}=t_{\vec{\delta}}^{ba}$ represents
the direction-dependent amplitude for an electron to hop from orbital $b$ to
orbital $a$ on site $i+\vec{\delta}$. ${\vec{A}}$ is the vector potential
and conductivity is calculated by linear response in ${\vec{A}}$ as usual.
The calculated band structure \cite{pickett,matheiss} is in fact well fit by
a $t_{\vec{\delta}}^{ab}$ which involves only nearest neighbor hopping which
is only non-zero for one particular linear combination of orbitals, and has
the magnitude $t_{0}\simeq 0.4$ eV. The particular linear combination
depends on the basis chosen for the $e_{g}$ doublet and on the direction of $%
\vec{\delta}$. If basis states $|{\bf x}^{2}-{\bf y}^{2}\rangle $ and $|3%
{\bf z}^{2}-r^{2}\rangle $ are chosen then for $\vec{\delta}=\hat{z}$ the
only non-zero element of $t$ is the one between $|3{\bf z}^{2}-r^{2}\rangle $
orbitals on the two sites. The analysis which follows does not depend in any
important way on the form chosen for $t_{\vec{\delta}}^{ab}$, but the
numbers of course do. In this model the quantity {\rm K}$={\rm K}(\omega
=\infty )$ defined in Eq.\ (\ref{K}) is given by 
\begin{equation}
{\rm K}={\frac{1}{{6N}}}_{sites}\sum_{%
{\scriptstyle i{\vec{\delta}} \atop \scriptstyle ab}%
}t_{\vec{\delta}}^{ab}\dot{\langle}d_{ia\alpha }^{\dag }d_{i+\delta b\alpha
}+h.c.\rangle .
\end{equation}
Here $\langle \rangle $ stands for a quantum and thermal expectation value
in the ensemble of eigenstates of ${\hat{H}}$. Note because ${\hat{H}}$
involves only a subset of all orbitals (in the present case, only the $%
e_{g}- $symmetry d$-$levels) the total optical oscillator strength in
transitions described by ${\hat{H}}$ is less than the full f-sum rule value N%
$e^{2}/m$ and indeed is given by the expectation value of an operator (in
the present case, the kinetic energy). This expectation value may depend on
temperature and interaction strength. The missing f-sum rule oscillator
strength comes from transitions between $e_{g}$ orbitals and other orbitals
not described by ${\hat{H}}$; these transitions will have oscillator
strengths which also depends on temperature and interaction. We are not
aware of theoretical results indicating which transitions are most important
in restoring the spectral weight. Interestingly, the data in Fig.~\ref{fig6}
suggest that in the manganites the $e_{g}-$O$_{2p}$ transitions are not the
ones primarily responsible for restoring the spectral weight. If $H_{INT}=0$%
, the non-interacting value {\rm K}$_{0}$ may be evaluated at $T=0$. For $%
x=0.3$ we find 
\begin{equation}
\text{{\rm K}}_{0}=0.46~t_{0}\cong 0.18\text{~eV}.  \label{kinetic}
\end{equation}
K$_{0}=0.18~$eV implies the effective carrier density {\rm N}$_{eff}=0.36.$

If $H_{INT}=0$ and a fully polarized ferromagnetic state is assumed for the
core spins then the free energy is given by extremizing the kinetic energy,
which is the operator whose expectation value yields K. Spin disorder or
inclusion of $H_{INT}$ will therefore change the wave function to one which
does not optimize the kinetic energy. K$_{0}$ is therefore an upper bound to
the optical spectral weight of the model specified by Eq.\ (\ref{hbar}) and
would be a reasonable estimate for the low-$T$ oscillator strength in the
manganite $e_{g}$ bands if interaction effects are not important. Some
caution is required because the Mn$-$Mn hopping amplitude is a very
sensitive function of the length and degree of buckling of the M$-$O$-$Mn
bond, so the value $t_{0}=0.4$ eV obtained from a band calculation for LaMnO$%
_{3}$ may not be exactly correct for the doped materials which have somewhat
smaller unit cell sizes and slightly different crystal structure.

In order to compare the observed oscillator strength to the band theory
estimate, we must identify the $e_{g}$ contribution to the absorption. In
the previous section the peak at $\omega \approx 4~$eV was identified as
coming from Mn$-$O transitions. From the $T=10$K curves in Fig.~\ref{fig2}
one sees that in the LSMO and LCMO samples the Mn$-$O transition dominates
the absorption for $\omega >2.7~$eV. For the NSMO sample the Mn$-$O
transition is not so clearly separated. For all three samples we estimate
the $e_{g}$ kinetic energy at $T=0$ by integrating the data up to $\omega
=2.7~$eV; this yields the values for {\rm K}$(2.7)$ shown in Table \ref
{table1}. The trend in kinetic energies is reasonable. LSMO has the highest $%
T_{c}$ and is the most metallic, consistent with its relatively large
optically determined kinetic energy, while NSMO has the lowest $T_{c}$ and
is the least metallic, and also has the smallest kinetic energy. All of the
observed kinetic energies are larger than those estimated from band theory.
The discrepancy could be due to an incorrect identification of the $e_{g}$
contributions to $\sigma$. If we cut off the integral at $2$ eV rather than $%
2.7$ eV the correspondence between data and band theory would be much more
reasonable. However, our analysis of the temperature dependence of the
spectral weight, which will be discussed below, leads us to believe that we
have correctly identified the $e_{g\text{ }}$transitions, and that our
analysis of the available band calculations has underestimated the kinetic
energy by up to 30\%; however in view of the uncertainties involved we
regard the correspondence between data and band theory as reasonable. The
analysis suggests that at low $T$ any renormalizations of the kinetic energy
due to electron-electron and electron-phonon interactions are small.

We now consider the temperature dependence of the spectral weight, which is
obviously large and provides further information on the relevant energy
scales. As the temperature is raised from zero the core spins cease to be
perfectly aligned, causing the kinetic energy to decrease. As explained in
Ref.~[\onlinecite{millisprl}] the decrease in bare electron kinetic energy
increases the effective electron-phonon coupling, causing a further decrease
in the $e_{g}$ kinetic energy. Additionally, the core spin disorder means
that optical transitions in which an $e_{g}$ electron goes from parallel to
the core spin on one site to antiparallel to the core spin on another
becomes possible. The final state energy in this ``antiparallel'' transition
is $\approx J_{H}S_{C\text{ }}$. In the $J_{H}\rightarrow \infty $ limit the
``antiparallel'' transitions may be neglected; in the actual materials $%
J_{H}S_{C\text{ }}$ is large enough that the ``parallel'' and
``antiparallel'' absorptions are well separated in energy and may be
discussed separately. At $T=0$, the core spins are fully aligned and only
``parallel'' transitions are possible; as $T$ is increased the spectral
weight in the parallel spin transitions decreases both because the total
kinetic energy decreases and because some of the remaining spectral weight
is transferred to the ``antiparallel'' transitions. A detailed theoretical
analysis of the temperature dependence of the spectral weight including
finite $J_{H}$ and realistic band structure is not yet available, but
results for $J_{H}=\infty $ and a simplified band structure are given in
Ref.~[\onlinecite{millisprb}]. For finite $J_{H}$ and more realistic band
structure statements can be made about the changes in spectral weights from $%
T=0$ to $T>T_{c}$, where the core spins may reasonably be expected to be
completely uncorrelated from site to site. Calculations using the
non-interacting model, Eq.\ (\ref{hbar}) with $H_{INT}=0$, show that if the
``parallel'' and ``antiparallel'' absorptions are well separated in energy
then as the temperature is raised from $T=0$ to $T>T_{c}$, the change in
spectral weight in the parallel spin transitions is given to good accuracy
by the results of the $J_{H}=\infty $ limit, while at $T>T_{c}$ the ratio of
``antiparallel'' to ``parallel'' spectral weights is approximately $%
2t_{0}/1.4J_{H}S_{c}\approx 1/3$ (using $t_{0}=0.4~$eV and $%
J_{H}S_{c}\approx 2~$eV, which is estimated as the difference between the
energies of the ``parallel'' and ``antiparallel'' absorption bands). We
believe that these conclusions, which have been verified in the
noninteracting limit, apply also to interacting models. We therefore analyze
the $T>T_{c}$ data by comparing the parallel spin spectral weight to the $%
J_{H}=\infty $ calculations and the ``antiparallel'' weight to $1/3$ the
parallel weight. The parallel-spin spectral weight is most conveniently
addressed by considering the change in oscillator strength between lowest $T$
and $T>T_{c}$. In double-exchange only models such as Eq.\ (\ref{hbar}) with 
${\rm H}_{INT}=0$, the oscillator strength decreases by $1/3$, i.e., $\Delta 
{\rm K}={\rm K}(T=0)-{\rm K}_{par}(T>T_{c})=1/3$ {\rm K}$(T=0)$. Models
involving the interplay of double-exchange and electron-phonon coupling can
produce a larger $\Delta {\rm K}/{\rm K}(T=0)$ because of a feedback effect:
decreasing {\rm K}$_{par}$ by increasing spin disorder increases the
effective electron-phonon coupling which decreases {\rm K}$_{par}$ still
further. In Ref.~[\onlinecite{millisprb}] weak, intermediate, and strong
electron-phonon couplings were distinguished. The intermediate coupling
regime was argued to be relevant to CMR; in this regime the phonon
renormalization of the kinetic energy is small at low $T$ (where the
behavior is metallic) but significant at high $T$ (where the interaction has
localized the electrons). This leads to $\Delta {\rm K}\cong \frac{1}{2}{\rm %
K}(T=0)$.

Table I presents the measured values of {\rm K}$_{par}(T>T_{c}),$ obtained
by integrating the observed conductivity up to $2.7$ eV. These values lead
to $\Delta {\rm K}\approx \frac{1}{2}{\rm K}(T=0)$, in good agreement with
theory. Further, the observed magnitude of $\Delta {\rm K}$ supports the
correctness of our estimate for {\rm K}$(T=0)$. If the band theory estimate K%
$_{0}=0.18$ eV were correct, then in LSMO, the observed change in spectral
weight would be 70\% of the total. Such a large change seems very unlikely.

Finally, we turn to the ``antiparallel'' absorption. Taking the 3 eV feature
in Fig.~\ref{fig6} to be the ``antiparallel'' transition and integrating the
difference in conductivity between the highest and lowest temperatures from
2.2 to 4 eV we find spectral weights {\rm K}$_{anti}$ given in Table \ref
{table1} for the three compounds. The parallel spectral weights {\rm K}$%
_{par}$ also shown in Table \ref{table1} are obtained as before by
integrating up to 2.7 eV. This leads to ratios of the ``antiparallel'' to
the ``parallel'' spin absorptions of about 1/3, in reasonable accord with
the theoretical estimates.

We now turn to the spin-wave dispersion which at small $q$ has the form $%
\omega _{magnon}=Dq^{2}$. The value of $D$ is of interest because in
double-exchange materials the spin stiffness is related to the amplitude for
conduction electrons to hop from one site to the next, and as we have seen
this is given by the conduction-band optical spectral weight. One therefore
expects a relation between the optical spectral weight and the spin wave
stiffness. To derive this relation ship more precisely, we consider the
energy cost $\Delta {\rm E}(q)$ of imposing a static distortion of
wavevector $q$. The long wavelength limit of this energy cost is related to
the spin wave stiffness by factors involving the quantum mechanical spin
dynamics. For double exchange materials the relationship is\cite{millisprl95}
\begin{equation}
D=\frac{1}{S^{*}}\frac{d^{2}}{dq^{2}}\Delta {\rm E}(q)\mid _{q=0}.
\end{equation}

Here $S^{*}=S_{c}+\frac{1}{2}(1-x)\cong 1.85$ is the magnetic moment per
site. To calculate $\Delta {\rm E}(q)$ we suppose that on each site the core
spin $S_{c}^{i}$ is characterized by polar angles ($\theta _{i},\phi _{i}$)
and that $\theta _{i}$ is small, so we are near the fully polarized $\theta
_{i}\equiv 0$ state. On each site we rotate the $e_{g}$\ electron basis so
that it is parallel to $\vec{S}_{ci}$. The rotation matrix which does this
is $R_{i}={\rm cos}{\frac{\theta _{i}}{{2}}}\hat{1}+i{\rm sin}{\frac{\theta
_{i}}{{2}}}{\vec{n}}\cdot {\vec{\sigma}}$ where ${\vec{n}}=(n_{x},n_{y})$
with $n_{x}={\rm cos}\phi _{i}$ and $n_{y}={\rm sin}\phi _{i}$. We now
calculate $E(\theta _{i})-E(\theta =0)$ by second order perturbation theory.
The perturbation Hamiltonian is 
\begin{equation}
{\hat{H}_{p}}=+{\frac{1}{{2}}}\sum_{%
{\scriptstyle {{{i\delta } \atop \scriptstyle {ab}}} \atop \scriptstyle \alpha \beta \gamma }%
}t_{\vec{\delta}}^{ab}\left[ ({\hat{1}}-R_{%
{i \atop \scriptstyle \gamma \alpha }%
}^{\dag }R_{%
{i+\delta  \atop \scriptstyle \gamma \beta }%
})d_{ia\alpha }^{\dag }d_{i+\delta b\alpha }+h.c.\right].
\end{equation}
Now $(1-R_{i}^{\dag }R_{i+\delta })$ has a term of order $\theta $, with
which we do second order perturbation theory, and a term of order $\theta
^{2}$, with which we do first order perturbation theory. For first order
perturbation theory we take the ground state expectation value of ${\hat{H}%
_{p}}$, obtaining $(\vec{m}_{i}=\theta _{i}{\vec{n}})$ 
\begin{equation}
E_{1}={\frac{{\rm K}}{{16}}}\sum_{i\delta }(m_{i}-m_{i+\delta })^{2}
\end{equation}
(note the term in ${\hat{1}}-R^{\dag }R$ proportional to $\sigma _{z}$
vanishes by symmetry). Now turn to second order perturbation theory
(neglected in some previous work\cite{millisprl95}). The second-order
perturbing Hamiltonian is 
\begin{equation}
{\hat{H}_{p}^{\prime }}=-{\frac{i}{{2}}}\sum_{%
{\scriptstyle {{{i\delta } \atop \scriptstyle {ab}}} \atop \scriptstyle \alpha \beta }%
}t_{\vec{\delta}}^{ab}\left[ ({\vec{m}}_{i}-{\vec{m}}_{i+\delta })\cdot {%
\vec{\sigma}}\right] (d_{ia\alpha }^{\dag }d_{i+\delta b\alpha }-h.c.).
\end{equation}
This is basically the current operator convolved with the gradient of the
magnetization. In the limit $J_{H}\rightarrow 0$, the up and down $e_{g}$\
bands become equally populated and for long wavelength spin fluctuations ${%
\hat{H}_{p}^{\prime }}$ becomes identical to the current operator; as $%
q\rightarrow 0$ the second order perturbation contribution cancels the first
order one just as do the paramagnetic and diamagnetic parts of the optical
conductivity and the stiffness vanishes. In the limit $J_{H}\rightarrow
\infty $ the contribution from ${\hat{H}_{p}^{\prime }}$ vanishes. We have
calculated the order $q^{2}$ term in the magnon dispersion to order 1/$J_{H}$
using Eq.\ (\ref{hbar}) with $H_{INT}=0$ and find 
\begin{equation}
D=\frac{{\rm K}a_{0}^{2}}{4S_{C}}\left[ 1-\frac{\alpha t^{2}}{J_{H}S_{c}{\rm %
K}}\right] ,  \label{magnon}
\end{equation}
with $\alpha =1.04$. A similar result was obtained by Furukawa using a
simpler electron dispersion and the dynamical mean field method. We expect a
very similar result for general interacting models, but with a possibly
different value of $\alpha$.

The band structure estimates $t=0.4$ eV and ${\rm K}=0.18$ eV, along with
the experimental value $J_{H}S_{C}=2$ eV gives $D=0.16~$eV$-${\AA }$^{2}$.
Use of the observed {\rm K} along with appropriately rescaled $t(t/t_{band}=%
{\rm K}_{observed}/{\rm K}_{band})$ gives $D=0.18~$eV$-${\AA }$^{2}$ for
NSMO and LCMO and $D=0.17~$eV$-${\AA }$^{2}$ for LSMO. The very weak
dependence on {\rm K} occurs because near ${\rm K}=0.5$ the variation in
first and second terms in Eq.\ (\ref{magnon}) almost exactly cancels.

Our estimates for $D$ are in excellent agreement with the experimental
values of $D=0.19~$eV$-${\AA }$^{2}$ for LSMO\cite{mcmartin} and $0.17$ eV$-$%
{\AA }$^{2}$ for both LCMO\cite{lynn} and NSMO.\cite{huang} Also, the weak
material dependence predicted by the theory is consistent with the
experimental finding. However, this close agreement may be somewhat
fortuitous since the parameters are not accurately known and the second
order correction to $D$ in Eq. (\ref{magnon}) is not very small compared
with the first order term. Nevertheless, these results suggest that other
effects that may be expected to affect the spin wave stiffness may not be
significant. One such effect proposed by Solovyev $et.al.$\cite{hamada} is
that in addition to the double-exchange ferromagnetic coupling there is a
weaker antiferromagnetic coupling between the core spins. This would act to
weaken the spin stiffness, and according to the calculations of Ref.~[%
\onlinecite{hamada}] its magnitude depends sensitively on the buckling of
the Mn-O-Mn bond. Also the electron-phonon coupling might affect the second
term in Eq.\ (\ref{magnon}). Therefore the issue warrants further
investigation.

\section{Conclusions}

We have measured and analyzed the optical conductivity of several manganite
materials. We have identified the physical origin of the various features in
the absorption spectrum and determined their variation with temperature and
material. The temperature dependence of the spectral weight agrees
quantitatively with theory and provides evidence in favor of the importance
of the electron-phonon coupling. The kinetic energy of the $e_{g}$ electrons
is found to be $10-20\%$ larger than predicted by band theory; in view of
the uncertainties in both calculation and measurement we believe this
constitutes good agreement. The relation between the optical spectral weight
and the spin wave stiffness was examined. Excellent agreement was found.

We would like to thank S. M. Bhagat, and S. E. Lofland for performing the
ferromagnetic resonance measurements on the samples prior to the optical
measurements and C. Kwon and C. Xiong for providing some of the samples. We
also thank A. G. Markeltz at NIST for performing the Terahertz measurements.
This work was supported in part by the NSF-MRSEC grant \# DMR-96-32521,
DMR-9705482 and NIST grant \# 70NANB5H0086.

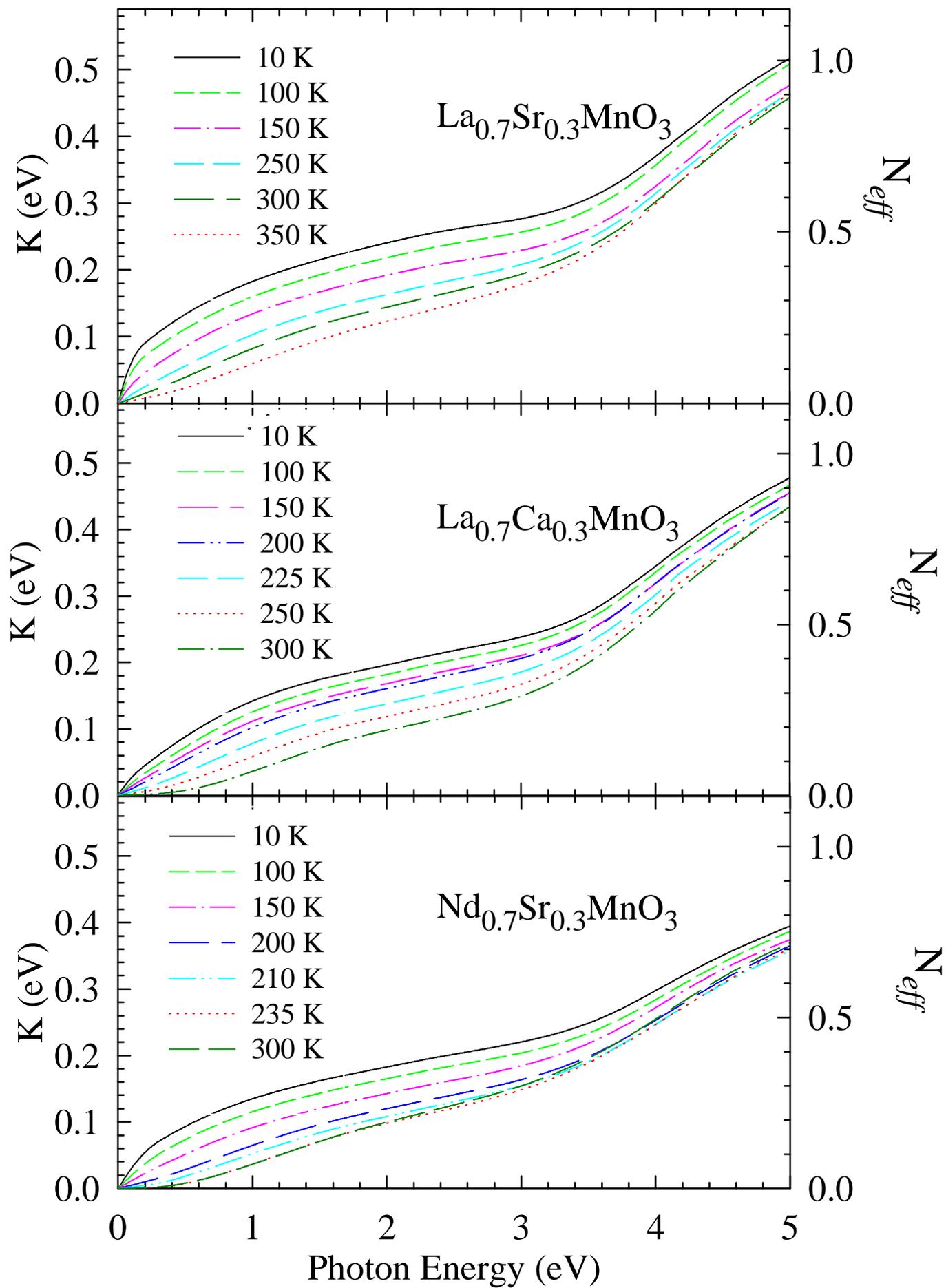
\begin{figure}[tbp]
\caption{Integral of the real part of the optical conductivity, 
\hbox{$\sigma_{1}(\omega)$}, as a funtion of photon energy. The results are
expressed in terms of kinetic energy and the carrrier density N$_{eff}$/f.u.}
\label{fig7}
\end{figure}

\begin{table}[tbp]
\caption{ Observed kinetic energies (in meV) and some kinetic energy ratios
for the three samples, obtained by integrating the observed conductivities
and using Eqs. 4.3 and 4.4. {\rm K}(2.7)$\equiv~${\rm K}$_{T=0}$($\omega=2.7$%
eV) is obtained by integrating the observed conductivity at lowest
temperature up to 2.7 eV; K$_{par}\equiv {\rm K}_{T>T_c}$($\omega=2.7$eV),
the spectral weight associated with parallel spin transitions, is obtained
by integrating the observed conductivity at the highest temperature up to
2.7 eV. K$_{anti}$, the conductivity assiociated with the antiparallel
transitions, is obtained by integrating the difference in conductivity
between highest and lowest temperatures from 2.2 to 4 eV.}
\label{table1}
\begin{tabular}{lccc}
& LSMO & LCMO & NSMO \\ 
\tableline K(2.7) & 260 & 220 & 200 \\ 
&  &  &  \\ 
K$_{anti}$ & 34 & 31 & 30 \\ 
K$_{par}$ & 125 & 100 & 100 \\ 
${\frac{{\rm K}_{anti}}{{\rm K}_{par}}}$ & 0.27 & 0.31 & 0.30 \\ 
${\frac{{\rm K}_{par}}{{{\rm K}(2.7)}}}$ & 0.48 & 0.45 & 0.50
\end{tabular}
\end{table}


\begin{references}
\bibitem[*]{byline}  Current address: Code 551, Goddard Space Flight Center,
Greenbelt, MD 20771.

\bibitem{kuster1}  R. M. Kusters, J. Singleton, D. A. Keen, R. Mcgreevy, and
W. Hayes, Physica B {\bf 155}, 362 (1989); R. Von Helmholt, J. Wecker, B.
Holzapfel, L. Schultz, and K. Samwer, Phys. Rev. Lett. {\bf 71}, 2331
(1993); K. I. Chahara, T. Ohno, M. Kasai, and Y. Kozono, Appl. Phys. Lett. 
{\bf 63}, 1990 (1993); S. Jin, T. H. Tiefel, M. McCormack, R. A. Fastnacht,
R. Ramesh, and L. H. Chen, Science {\bf 264}, 413 (1994); M. McCormack, S.
Jin, T. H. Tiefel, R. M. Fleming, J. M. Phillips, and R. Ramesh, Appl. Phys.
Lett. {\bf 64}, 3407 (1994); H. L. Ju, C. Kwon, Q. Li, R. L. Greene, and T.
Venkatesan, Appl. Phys. Lett. {\bf 65}, 2109 (1994).

\bibitem{zener}  C. Zener, Phys. Rev. B {\bf 82}, 403 (1951); P. W. Anderson
and H. Hasegawa, Phys. Rev. {\bf 100}, 675 (1955); P.-G. de Gennes, Phys.
Rev. {\bf 118}, 141 (1960).

\bibitem{millisprl95}  A. J. Millis, P. B. Littlewood, and Boris I.
Shraiman, Phys. Rev. Lett. {\bf 74}, 5144 (1995).

\bibitem{millisprl}  A. J. Millis, R. Mueller, and Boris I. Shraiman, Phys.
Rev. Lett. {\bf 77}, 175 (1996).

\bibitem{millisprb}  A. J. Millis, R. Mueller, and Boris I. Shraiman, Phys.
Rev. B {\bf 54}, 5405 (1996).

\bibitem{rother}  H. Roder, Jun Zang, and A. R. Bishop, Phys. Rev. Lett. 
{\bf 76}, 1356 (1996).

\bibitem{jung}  J. H. Jung, K. H. Kim, D. J. Eom, T. W. Noh, E. J. Choi,
Jaejun Yu, Y. S. Kwon, and Y. Chung, preprint.

\bibitem{hamada}  I. Solovyev, N. Hamada, and K. Terakura, Phys. Rev. B{\bf %
53}, 7158 (1996).

\bibitem{okimoto}  Y. Okimoto, T. Katsufuji, T. Ishikawa, T. Arima, and Y.
Tokura, Phys. Rev. B{\bf 55}, 4206 (1997).

\bibitem{tokura}  Y. Okimoto, T. Katsufuji, T. Ishikawa, A. Urushibara, T.
Arima, and Y. Tokura, Phys. Rev. Lett. {\bf 75}, 109 (1995).

\bibitem{arima}  T. Arima, Y. Tokura, and J.B. Torrance, Phys. Rev. B{\bf 48}%
, 17006 (1993).

\bibitem{kim}  K. H. Kim, J. Y. Gu, E. J. Choi, G. W. Park, and T. W. Noh,
Phys. Rev. Lett. {\bf 77}, 1877 (1996).

\bibitem{billinge}  S. J. L. Billinge, R. G. Difrancesco, G. H. Kwei, J. J.
Neumeier, and J. D. Thompson, Phys. Rev. Lett. {\bf 77}, 715 (1996).

\bibitem{asamitsu}  A. Asamitsu, Y. Moritomo, Y. Tomioka, T. Arima, and Y
Tokura, Nature (London) {\bf 373}, 407 (1995).

\bibitem{jaime}  M. Jaime, M. B. Salamon, K. Pettit, M. Rubinstein, R. E.
Treece, J. S. Horwitz, and D. B. Chrisey, Appl. Phys. Lett. {\bf 68}, 1576
(1996).

\bibitem{mahan}  G. D. Mahan, {\it Many-particle Physics} (Plenum Press, New
York, 1981), Chap 6.

\bibitem{emin}  David Emin, Phys. Rev. B{\bf 48}, 13691, (1993).

\bibitem{simon}  S. Kaplan, M. Quijada, H. D. Drew, D. B. Tanner, G. C.
Xiong, R. Ramesh, C. Kwon, and T. Venkatesan, Phys. Rev. Lett. {\bf 77},
2081, (1996).

\bibitem{2}  G. C. Xiong, Q. Li, H. L. Ju, S. N. Mao, L. Senapati, X. X. Xi,
R. L. Greene, and T. Venkatesan, Appl. Phys. Lett. {\bf 66}, 1427 (1995).

\bibitem{lofland}  S. E. Lofland, S. M. Bhagat, H. L. G. C. Xiong, T
Venkatesan, R. L. Greene, and S. Tyagi, J. Appl. Phys. {\bf 79}, 5166 (1996).

\bibitem{sam}  S. Lofland and S. Bhagat, Private communication.

\bibitem{coey}  J. M. D. Coey, M. Viret, L. Ranno, and K. Ounadjela, Phys.
Rev. Lett. {\bf 75}, 3910 (1995).

\bibitem{rama}  T. V. Ramakrishnan, {\it IX Trieste Workshop on Open
Problems in Strongly Correlated Systems}, Trieste (1997).

\bibitem{andrea}  A. G. Markeltz, Private communication.

\bibitem{heavens}  O. S. Heavens,{\it Optical Properties of Thin Solid Films}
(Dover Publications Inc.,1991), Chapter 4.

\bibitem{wooten}  F. Wooten, {\it Optical Properties of Solids} (Academic
press, New York, 1972), Chapter 6.

\bibitem{zhang}  Z. M. Zhang, B. I. Choi, M. I. Flik, and A. C. Anderson, J.
Opt. Soc. Am. B {\bf 11}, 2252 (1994).

\bibitem{korean98}  J. H. Jung, K. H. Kim, T. W. Noh, E. J. Choi, and Jaejun
Yu, Preprint.

\bibitem{Andywillget}  P. G. Radaelli, D. E. Cox, M. Marecio, and S-W
Cheong, Phys. Rev. B{\bf 55}, 3015 (1997).

\bibitem{chainani}  A. Chainani, M. Mathew, and D. D. Sarma, Phys. Rev. B 
{\bf 47}, 15397 (1993).

\bibitem{houstonPRL}  L. Sheng, D. Y. Xing, D. N. Sheng, and C. S. Ting,
Phys. Rev. Lett. {\bf 79}, 1714 (1997).

\bibitem{jrcat}  Y Tokura, A. Asamitsu, Y. Tomioka, H. Kuwahara, T. Okuda,
Y. Okimoto, T. Ishikawa, T. Katsufuji, and E. Saitoh, {\it Workshop on Phase
Control of Colossal Magnetoresistive Oxides}, Tsukuba (1997).

\bibitem{booth98}  C. H. Booth, F. Bridges, G. H. Kwei, J. M. Lawrence, A.
T. Carnelia, and J. J. Neumeier, Phys. Rev. Lett. {\bf 80}, 853 (1998).

\bibitem{pickett}  W. E. Pickett and D. Singh, Phys. Rev. B {\bf 53}, 1146
(1996).

\bibitem{matheiss}  L. F. Matheiss, unpublished.

\bibitem{lynn}  J. W. Lynn, R. W. Erwin, J. A. Borchers, Q. Huang, A.
Santoro, J-L. Peng, and Z. Y. Li, Phys. Rev. Lett. {\bf 76}, 4046 (1996).

\bibitem{mcmartin}  M. C. Martin, G. Shirane, Y. Endoh, K. Hirota, Y.
Moritomo, and Y. Tokura, Phys. Rev. B {\bf 53}, 14285 (1996).

\bibitem{huang}  H. Y. Hwang, private communication.
\end{references}
\end{document}